\documentclass{raa}

\usepackage{graphicx,times}             
\input{epsf.sty}                        
\input{psfig.sty}                       

\begin{document}

\title{Periodic Radio Variability in NRAO 530: Phase Dispersion Minimization Analysis}

   \volnopage{Vol.0 (200x) No.0, 000--000}      
   \setcounter{page}{1}          
\author{Jun-Chao Lu
      \inst{1}
   \and Jun-Yi Wang
      \inst{2,1}
   \and Tao An
      \inst{3,4}
   \and Ji-Ming Lin
      \inst{1}
   \and Hong-Bing Qiu
      \inst{1}
   }

   \institute{Key Laboratory of Cognitive Radio $\&$ Information Processing, the Ministry of Education, Guilin University of Electronic Technology, Guilin 541004, China; {\it ljckiller@163.com}, {\it wangjy@guet.edu.cn}, {\it linjm@guet.edu.cn}\\
       \and
        School of Mathematics and Computing Science, Xiangtan University, Hunan 411105, China;
        \and
             Shanghai Astronomical Observatory, Chinese Academy of Sciences, Shanghai 200030, China;  {\it antao@shao.ac.cn} 
   \and 
Key Laboratory of Radio Astronomy, Chinese Academy of Sciences, China. \\
   }

   \date{Received~~2009 month day; accepted~~2009~~month day}

\abstract{In this paper, a periodicity analysis of the radio light curves of the blazar NRAO~530 at 14.5, 8.0, and 4.8 GHz is presented employing an improved Phase Dispersion Minimization (PDM) technique. The result, which shows two persistent periodic components of $ \sim 6$ and $ \sim 10$ years at all three frequencies, is consistent with the results obtained with the Lomb-Scargle periodogram and weighted wavelet Z-transform algorithms.
The reliability of the derived periodicities is confirmed by the Monte Carlo numerical simulations which show a high statistical confidence.
(Quasi-)Periodic fluctuations of the radio luminosity of NRAO~530 might be associated with the oscillations of the accretion disk triggered by hydrodynamic instabilities of the accreted flow.
}
   \authorrunning{J.-C. Lu et al.}            
   \titlerunning{PDM periodicity analysis of NRAO~530}  

   \maketitle
%
\section{Introduction}           
\label{sect:intro}
Active Galactic Nuclei (AGNs) are the most luminous objects in the Universe, and understanding the AGN phenomenon is an important issue in modern astrophysics.
The energy production process of AGNs is one of the fundamental problems to be solved in extragalactic astrophysical studies. Currently direct imaging of the central engines of AGNs is still constrained by insufficient resolution. However, the studies of variability of AGNs and the radiation spectrum are not limited by two-dimensional mapping resolution.
Blazars constitute a sub-class of the large family of AGNs, which are characterized by rapid and violent time variability across almost the whole spectrum, high polarization, compact relativistic jets beaming toward the observers.
An amount of blazars have been found to display periodic or quasi-periodic variabilities in the continuum emission (e.g., OJ~287: Sillanpaa et al. 1988).
Periodicity in light curves of blazars is intimately associated with the dynamical progress occurring in the central engine of the active nuclei. One can obtain rich information of the internal structure (as well as the variation) and the energy production process from periodicity analysis (Ulrich, Maraschi \& Urry, 1997). But the complexity of variability origins, the interplay of different variability modes, and the lack of time coverage of the monitoring data make the identification of genuine periods difficult (Fan et al. 2001). Therefore, a well-sampled light curve covering a sufficiently long time span and an accurate and efficient analysis technique are of particular importance for identifying the periodicity of blazars with time scale of a few years.

Traditional numerical methods used for searching for and identifying periodic fluctuations in time series are based on Fourier transform (Lomb 1976; Scargle 1981, 1982, 1989). It works well when the signals have simply sinusoidal shapes and the periodicity parameters (such as characteristic frequency, phase) remain constant over the whole time span. However the real blazar light curve is often a mixture of periodic signals of different time scales, and is always blended with intermittent or short-lifetime flares which can not be expressed with simple sinusoidal functions in time domain. In addition, irregular gaps in the monitoring data, non-sinusoidal variations, and other problems would make Fourier techniques unusable and yield misleading results. In contrast, Phase Dispersion Minimization (PDM) is a good algorithm in periodicity analysis of the time series with non-sinusoidal oscillation functions and irregular sampling (Stellingwerf 1978; Linnell Nemec \& Nemec 1985; Schwarzenberg-Czerny 1989, 1997). Compared to the Fourier-based algorithm, the PDM has the advantages in computing speed, implementation efficiency, and statistical confidence, especially in distinguishing true periodic signals from noises.

In this paper, we search for periodicity in the radio light curves of the blazar NRAO~530, making use of the PDM technique. The blazar NRAO~530 (also named PKS~1730-130) is a well-known Optically Violently Variable (OVV) quasar at a redshift of 0.902 (Junkkarinen 1984). It exhibits prominent variability across the whole electromagnetic spectrum covering the radio, optical, UltraViolet, X-ray and $\gamma$-ray bands (Hong et al. 2008). NRAO~530 is one of the few blazars showing persistent quasi-periodicity on time scale of a few years (Hovatta et al. 2008; An et al. 2012) and is a good candidate for investigating long-term periodic variabilities and the associated physical processes.
The remaining text is organized as follows: Section 2 describes the radio monitoring data of NRAO~530 used for this analysis; Section 3 introduces the basic principle of the PDM algorithm, and presents the periodicity analysis results, and tests the statistical confidence of the periods using Monte Carlo simulations; a comparison of the results from the PDM with other techniques are presented in Section 4; a summary and conclusion is given in Section 5.
\section{The Data}
\label{sect:data}
\begin{figure}[h]
   \centering
   \includegraphics[width=0.8\textwidth]{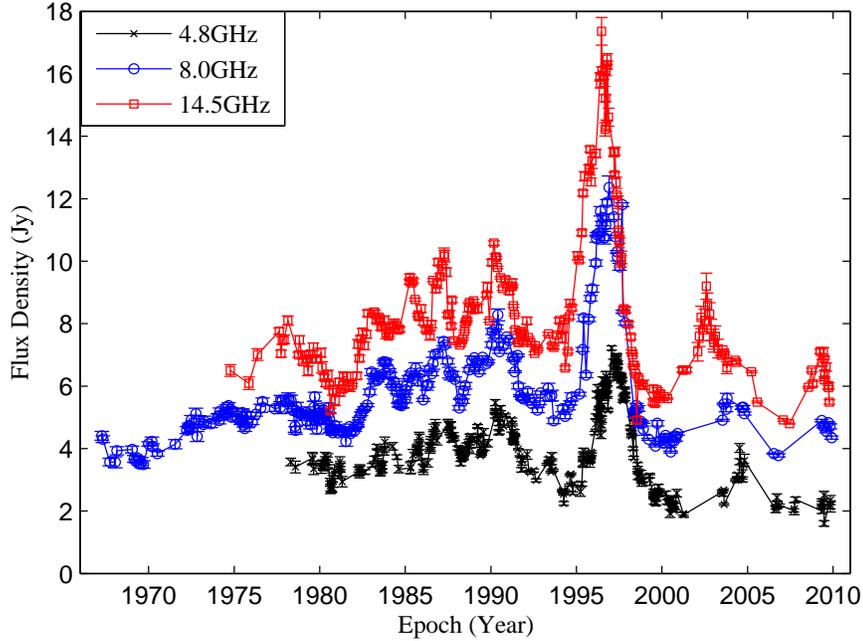}
   \caption{{\small Light curves of NRAO~530 at three frequencies of 14.5 (red square), 8.0 (blue circle) and 4.8~GHz (black cross). In order to avoid the overlapping of data points, the 4.8-GHz light curve has been moved downward by 2~Jy, and 14.5-GHz flux density has been moved upward by 2~Jy. A major outburst is clearly seen in 1996-1997.}}
   \label{Fig:demo1}
   \end{figure}
The total flux density data of NRAO~530 are observed with the 26-meter radio telescope of the University of Michigan Radio Astronomical Observatory (Aller et al. 1985) at three frequencies of 14.5, 8.0 and 4.8~GHz. The earliest epoch of the observational data is on 1 April 1978 (4.8~GHz), 19 April 1967 (8.0~GHz) and 10 October  1974 (14.5~GHz), respectively. The latest epoch is in November 2009 for all three frequencies. The maximum length of the time span is about 35 years, which is sufficiently long to search for any quasi-periodicity in the light curves with a few complete cycles. Considering that the time scale of radio outbursts is at the magnitude from a few months to years, the bi-weekly time sampling is good enough for performing accurate periodicity analysis.
\section{Periodic Variability Of NRAO~530 using PDM Method}
\subsection{The Basic Principle of the PDM}
 The PDM is a widely used numerical technique searching for and identifying periodic fluctuations of the luminosity of variable stars and other astronomical objects (Stellingwerf 1978). In the PDM algorithm, the time series data $\{(x_i, t_i)\}$  are divided into a number of phase bins, where $x_i$ is amplitude of the $i$th data point at the observing time $t_i$. Let $N$ be the total number of the data points and $M$ be the number of the phase bins, and let $x_{kj}$ represent the $k$th data point in the $j$th phase bin. A proper $M$ is chosen according to the trial period, so that for a given trial period, the data points in each phase bin have similar phases. The phase of the $i$th data point is denoted as ${\phi _i} = {t_i}/\Pi  - [{t_i}/\Pi ],\; i = 1,\cdots,N,\;\;{\phi _i} \in [0,1]$. The bins can overlap to improve the phase coverage, if necessary. The variance $\sigma^2$ of the whole data set is expressed as
\begin{equation}
\sigma^2 = \frac{\sum(x_i - \bar{x})^2}{N-1}, \;\;\; (i=1,\cdots,N).
\end{equation}
The sample variance within each bin is defined in the same way as $s_j^2 =  \frac{\sum(x_{kj} - \bar{x}_j)^2}{n_j-1},\; (j=1, \cdots,M)$, where $n_j$ is the number of data points in the $j$th phase bin. The overall variance for all phase bins $s^2$ is expressed as
\begin{equation}
s^2 = \frac{\sum(n_j - 1) s_j^2}{\sum{n_j} - M}, \;\;\; (j=1,\cdots,M).
\end{equation}
The PDM picks the mean in each phase bin $\bar{x}_j$ to construct a mean light curve which is used as a reference of the time variation, then computes the scatter of the data points with respect to the mean light curve. This scatter is quantitatively evaluated by a statistical quantity $\Theta = {s^2}/{\sigma^2}$ (the ratio of the phase bin variance to the total data variance). For a true period, $\Theta$ will be arbitrarily small; for a false period, $\Theta$ will be approximately unity. The PDM searches for a series of testing periods, and seeks for the periods at which the scatter of the observed light curve from the mean light curve is minimized. A plot of the ratios ($\Theta$) versus trial periods provides an indicator of the best candidates for periodic components. The real periods are identified as the ones corresponding to local minimum values of $\Theta$ in the plot. The significance of the detected PDM periods has been improved from the original Fisher-Snedecor distribution (Stellingwerf 1978) to a $\beta$ distribution (Linnell Nemec \& Nemec 1985; Schwarzenberg-Czerny 1989, 1997). Quantitative evaluation of the significance can be obtained using the Monte Carlo simulations.
\subsection{Testing the PDM Using a Simulated Multiple-Period Signal}
In order to verify the reliability and applicability of the PDM approach, we adopt a simulated time series with a function of  $y = \sin (\frac{\pi }{5}t) + 1.5\sin (\frac{{2\pi }}{3}t)$ to perform the PDM analysis. It is a combination of two sinusoidal functions with different periods (10 and 6 year) and different amplitudes (1.0 and 1.5). This simulated signal is sampled in accordance with the same time interval of NRAO~530 observing data to construct discrete time series. And a discrete random noise signal with the amplitude range of [0,0.1] is also added to the time series. The time coverage of the simulated time series is from 1975 to 2010, same with observing data. The simulated signal is shown in Fig. 2. The variation of the resulting $\Theta$s as a function of trial periods is shown in Fig. 3. The plot highlights the two major periods between 0--0.25 ($year^{- 1}$), and the $\Theta$ fluctuates between 0.7--0.9 on high frequency (shorter period) part of the plot, which is not shown. The plot clearly displays two periods of 9.87 and 6.03 years, corresponding to the first and secondary minimum of $\Theta$, respectively. The results agree with the initial definition of 10 and 6 year periods in the signal function. It demonstrates that PDM is not only able to identify periodic fluctuations, but also to detect multiple periodic components.
\begin{figure}[h]
  \begin{minipage}[t]{0.495\linewidth}
  \centering
   \includegraphics[width=70mm,height=45mm]{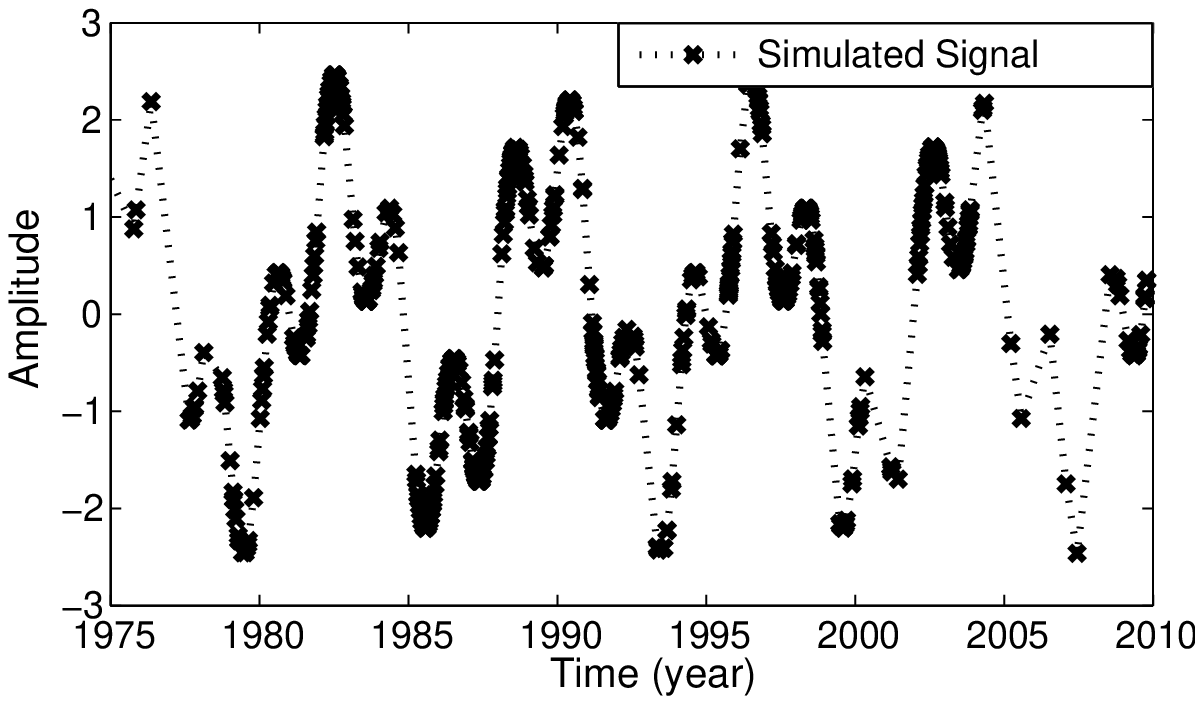}
   \caption{{\small Simulated time series with periods of 10 and 6 years with different amplitudes. The time span is used the same with the observing data for easy comparison.}}
  \end{minipage}%
  \begin{minipage}[t]{0.495\textwidth}
  \centering
   \includegraphics[width=70mm,height=45mm]{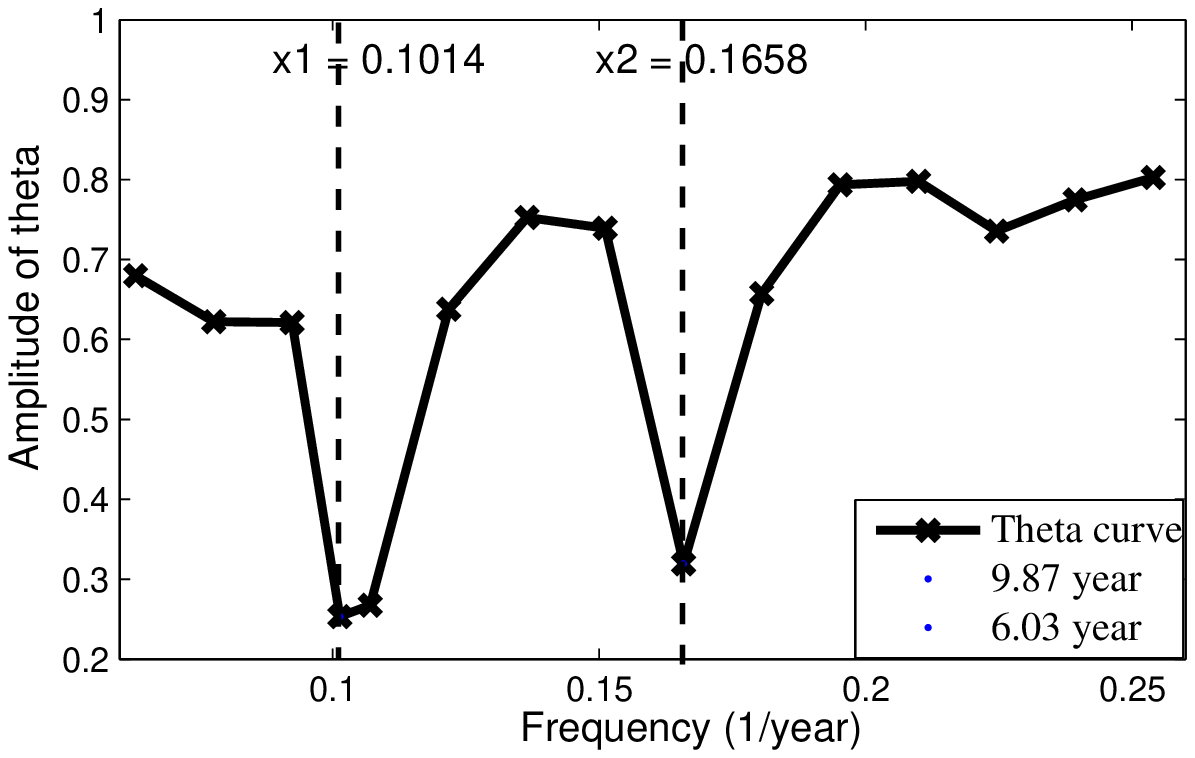}
  \caption{{\small The variation of $\Theta$ with trial periods. The periods are recognized as the ones at which $\Theta$ shows a local minimum.}}
  \end{minipage}%
  \label{Fig:fig2}
\end{figure}
\subsection{PDM Periodogram of NRAO~530}
We would conduct the PDM analysis in two steps. Firstly, the data were divided into a number of segmentations in a 'rough-cut' way. The algorithm performed abroad search for periodic signals with the data gaps removed. This resulted in a group of candidates for the periodic frequencies. Secondly, the segmented data were computed by PDM algorithm to scan the trial frequencies near the candidates obtained in the first step. The gaps among the observing data were included in the calculation to obtain more accurate periods. This second scan ('fine scan') gives a fine resolution of the detected characteristic frequencies. The running of the algorithm in two steps greatly saves the computing time, and increase the efficiency.

The periodicity search with the PDM method was made across a time baseline of 3 decades. Considering that the total time span of the monitoring data is about 30-35 years, it is expected to detect periodic signals with a few complete cycles during the observed time. Since the PDM approach used in this paper is aimed to identifying long-period variabilities, we set the trial frequencies in the range from 0.067 to 1 ($year^{- 1}$), which is equivalent to a trial period range of 1--15 years. According to the selected parameters and the procedure of the PDM approach described above, we obtained the $\Theta$-Frequency relation at three observing bands. Fig. 4 shows the $\Theta$ statistics versus the testing frequency for the radio data of NRAO~530. At all three frequencies, the primary and secondary minima are detected at similar positions, {\it i.e.}, $\sim$0.0960 ($year^{- 1}$) and $\sim$0.1695 ($year^{- 1}$), which correspond to the periodicity time scales of 10.4 and 5.9 years, respectively. These two minima are less prominent at 4.8~GHz (Top panel: Fig. 4) when compared with those at 8.0~GHz (Middle panel: Fig. 4) and 14.5~GHz (Bottom panel: Fig. 4). The reason might be associated with the increasing opacity at lower observing frequency that tends to smooth the variabilities. In addition to these two significant minima, the diagrams also show some other local minima but with less distinction.
At 14.5~GHz, a minimal $\Theta_{min}$ $\sim$0.75 corresponds to a period of $\sim$3.2 year (frequency = 0.31 $year^{- 1}$).
At 8.0~GHz, a local minimum $\Theta_{min}\sim0.7$ corresponds to a period of $\sim$4.9 year, which is also seen at 4.8~GHz at a similar period $\sim5$ year.
At 4.8~GHz, there are some hints of shorter periods at $\sim$2.5 year and $\sim$1.8 year.
\begin{figure}[h]
  \begin{minipage}[t]{0.495\textwidth}
  \centering
   \includegraphics[width=75mm,height=65mm]{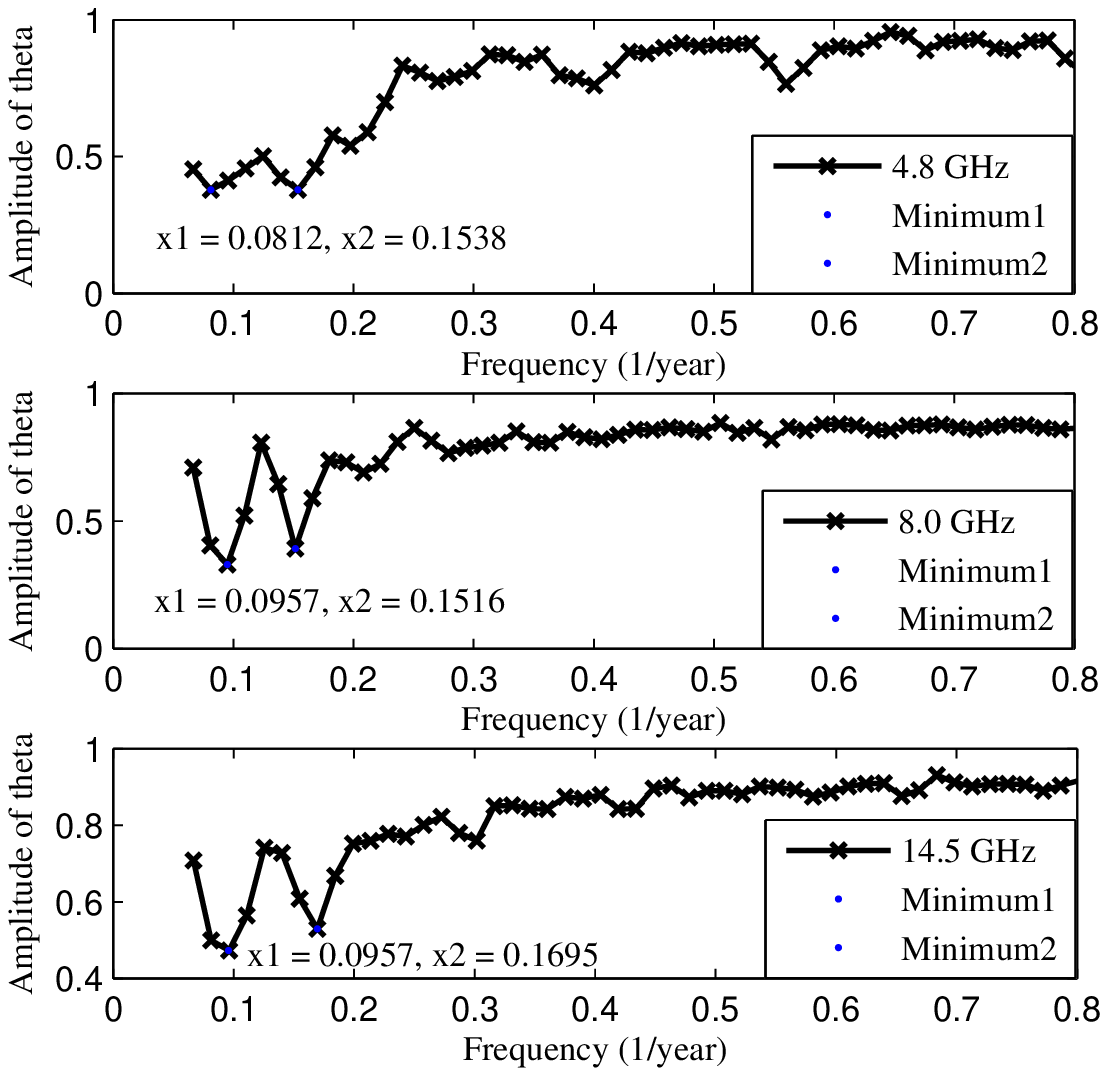}
  \caption{{\small $\Theta$-frequency diagrams of the NRAO~530: 4.8~GHz (Top), 8.0~GHz (Middle) and 14.5~GHz (Bottom). The oscillations are aliases caused by the data gaps.}}
  \end{minipage}%
  \label{Fig:fig3}
    \begin{minipage}[t]{0.495\linewidth}
  \centering
   \includegraphics[width=75mm,height=65mm]{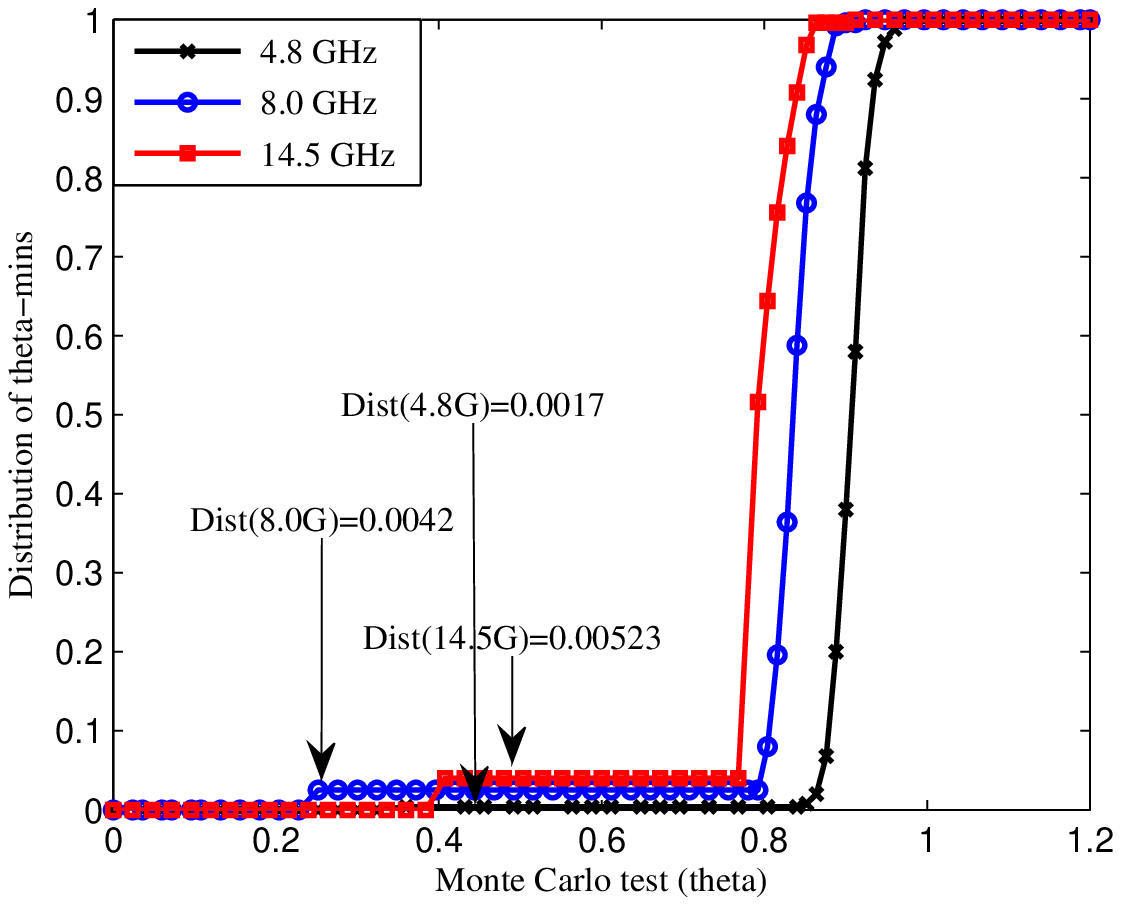}
   \caption{{\small Statistic significance test using Monte Carlo simulations. Three curves represent the $\Theta_{min}$ distribution of the 10-year periodicity at three radio frequencies of 14.5, 8.0 and 4.8~GHz.} }
   \label{Fig:fig4}
  \end{minipage}%
\end{figure}
In order to test the statistic significance of the derived periodicities, we performed Monte Carlo simulations (Nemec \& Nemec 1985) to the most prominent periods of $\sim$6 and $\sim$10 years. When employing the numerical simulations, the original observing data (the epochs and amplitudes) were used, but the data sequence were randomized. A normal PDM analysis was made again for these randomized data sets using the same parameters adopted for the actual data analysis. The integral of this distribution of $\Theta_{min}$ represents an estimate of the probability of $\Theta_{min}$ distribution could be due to pure noise.
The Monte Carlo simulation was repeated 500 times until obtaining a full distribution of noise-generated $\Theta_{min}$ distribution. This distribution was then used to estimate the statistic significance of the testing periods. Fig. 5 shows an example of the Monte-Carlo simulation results of 10-year periodicity after 500 times of trials. The probability for generating a 10-year period from a pure noise signal is 0.00523 (14.5~GHz), 0.0042 (8.0~GHz) and 0.0017 (4.8~GHz). That converts to a probability for a genuine 10-year period, 99.48\% (14.5~GHz), 99.58\% (8.0~GHz) and 99.83\% (4.8~GHz).
The characteristic periods computed by the PDM approach and the statistic significance derived from the Monte Carlo simulations are tabulated in Table 1. Note that only the two self-consistent periods which appear at all three observing bands are listed.
The confidence of the 6-year period is 99.52\%, 99.46\%, 99.97\% for 14.5, 8.0 and 4.8~GHz, respectively.
\begin{table}[h]
\begin{center}
\caption{Periodicity analysis results of NRAO~530 and Monte Carlo simulations.}
\label{Tab:publ-works}
\begin{tabular}[9cm]{cccccc} \hline
$\nu$ &NO.  & Freq.   &  P    & $\Theta_{min}$       & Signif \\
(GHz) &    & ($year^{- 1}$)  & (year)  &          & Monte Carlo \\ \hline
14.5  & 1  & 0.0960  & 10.4  & 0.4730   & 0.00523        \\
      & 2  & 0.1695  &  5.9  & 0.5293   & 0.00479          \\
8.0   & 1  & 0.0949  & 10.5  & 0.3289   & 0.0042            \\
      & 2  & 0.1515  &  6.6  & 0.3915   & 0.0054                    \\
4.8   & 1  & 0.0812  & 11.3  & 0.3778   & 0.00017          \\
      & 2  & 0.1537  &  6.5  & 0.3786   & 0.00035          \\
\hline
\end{tabular}
\end{center}
\end{table}
\section{Comparison with results from other methods}
In order to verify the results derived from the PDM, we also analyzed the variability data of NRAO~530 making use of the Fourier-transform method (Lomb-Scargle periodgram: Lomb 1976; Scargle 1982; Scargle 1989) and Wavelet transform method (Grossmann et al. 1989; Foster 1996). The results obtained from the Lomb-Scargle and Wavelet methods were discussed in a parallel paper (An et al. 2012).

Lomb-Scargle algorithm (LS) is a variation of the Discrete Fourier Transform (DFT), in which a time series is decomposed into a linear combination of sinusoidal functions. The basis of the algorithm is to transform the data with sinusoidal functions from the time domain to the frequency domain. Scargle (1981, 1982) has developed a set of formula for the transform coefficients that have similar forms with the DFTs when conducting evenly spaced time series. In addition, an adjustment in calculation of the transform coefficients has been made to make the transform insensitive to time shifts. The power of the LS periodogram is defined as the inverse of the variance of the fit as a function of the trial frequency. The best-fitted period is identified as the one giving the maximum power. The Lomb-Scargle periodogram detected a number of local maxima in the power-period plot, among which the highest two peaks corresponding to 6.2- and 10-year periods (An et al. 2012). Other lower power peaks indicated shorter-timescale periods of $\sim$5, $\sim$4 and $\sim$3 years.

Wavelet transform is another method widely used for time variability analysis. The peculiar character of the wavelet method is the localization property in both time and frequency domains. This localization property allows the wavelet functions to have limited durations, thus enables the wavelet transform to capture transient flares or short-lived periodic signals.
The basic procedure of the wavelet transform analysis is first to construct a mother wavelet, then to convolve the mother wavelet with a set of trial functions. The wavelet functions, in principle, might be of any shapes, therefore the wavelet transform may naturally identify the non-sinusoidal periodic signals.
An et al. (2012) used the Weighted Wavelet $Z$-transform (WWZ: Foster 1996) to search for the periodicity of NRAO~530.
The powers of the wavelet transform, {\it WWZ}, are computed as a function of time ($t$) and oscillation frequency ($\omega$).
Two persistent periodicities of $\sim$6-year and $\sim$10-year were detected from the {\it WWZ}. This is consistent with the results from the present PDM analysis. In addition, some other periodic signals with timescales of $\sim$5 year, $\sim$4 year, $\sim$3 year and $\sim$1.8 year were detected but with relatively lower powers. These shorter-timescale periodicities are also detected in the PDM diagrams (Fig. 4).
Table 2 compares the major periods detected from the NRAO~530 data using the PDM (the present paper), LS and {\it WWZ} methods (An et al. 2012).

\begin{table}[h]
\begin{center}
\caption{Periodicity Analysis of the Radio Light Curves Of NRAO~530 with the PDM, LS and WWZ.}
\label{Tab:publ-works}
 \begin{tabular}{c|ccc|ccc|ccc}
\hline
Method        & \multicolumn{3}{c|}{PDM} &  \multicolumn{3}{c|}{LS} &  \multicolumn{3}{c}{WWZ} \\
$\nu$(GHz)    & 4.8 & 8.0 & 14.5 & 4.8 & 8.0 & 14.5 & 4.8 & 8.0 & 14.5  \\
\hline
Period 1(Year)&11.3 & 10.5& 10.4 & 9.5 & 10.4& 9.9  & 9.7 & 10.1& 9.8   \\
Period 2(Year)& 6.5 & 6.6 & 5.9  & 6.3 & 6.3 & 6.2  & 6.0 & 6.0 & 5.9   \\
\hline
\end{tabular}
\end{center}
\end{table}

\section{Summary and Conclusion}
\label{sect:conclusion}

The PDM method was employed to search for the periodicity from the radio data of NRAO~530. The main results of the present paper can be summarized as :
\begin{itemize}
\item Multiple periods of different timescales have been identified.
The $\Theta$-period diagrams display two distinctive minima that have comparable magnitudes. The corresponding periods are $\sim$10.5 year and $\sim$6.5 year. The self-consistency of these two major periodic signals at all three observing frequencies suggests that the periodic fluctuations of the radio luminosity are associated with persistent dynamic processes. Other periodicities with shorter timescales and less distinction have also been detected.
\item Statistic significance of two prominent periods (10- and 6-year)  has been tested using the Monte Carlo technique.  The numerical simulations showed high confidences for both periodic signals at all three observing bands, confirming the reality of the PDM results.
\item Excellent consistency was demonstrated from the comparison of the periodicity results derived from the PDM, LS and WWZ methods, proving the validity of these numerical techniques in the AGN variability studies.
\end{itemize}
Different methods arrive at a consistent conclusion that there are multiple periodicities in the radio light curves of NRAO~530. Moreover there periodicities are found to have a harmonic frequency relation, $n\times f_1$, where $n=1,2,3,...$ (An et al. 2012). However, the fundamental oscillation frequency $f_1\approx \frac{1}{20}$~year$^{-1}$ is not detected over the total 35-year time span. We argue that the two prominent periods ($\sim$10 year and $\sim$6 year) may represent the $2nd$ and $3rd$ harmonics. The higher-order harmonics are suppressed during the active periods of radio outbursts. The multiple harmonics of the periods seem not a unique property of NRAO~530. This phenomenon has also been observed in other blazars, such as 3C279 (Liu et al. 2006), 1510$-$089 (Xie et al. 2008). Physical processes related to precessing jets driven by orbiting black hole binaries or warped accretion disks are not satisfactory to explain the present observations. Whereas $p$-mode oscillations triggered by disk instabilities are found to provide reasonable interpretations of the multiple periodicities and their harmonic relations (An et al. 2012). Variability studies of a large sample of blazars are necessary to verify whether the present analysis results and implications are universal.

\begin{acknowledgements}
This work has been partly supported by the  National Basic Research Program of China (2009CB24900), Guangxi Natural Science Foundation (0991018Z), and the Science and Technology Commission of Shanghai Municipality (06DZ22101).
Tao An has been supported by a grant from the Chinese Academy of Sciences and a visitor grant from the The Netherlands Science Foundation.
The University of Michigan Radio Astronomy Observatory is supported by funds from the NSF, NASA, and the University of Michigan.
\end{acknowledgements}

\label{lastpage}

\end{document}